\documentclass[a4paper,fleqn,usenatbib]{mnras}

\usepackage{newtxtext,newtxmath}

\usepackage[T1]{fontenc}
\usepackage{ae,aecompl}

\usepackage{graphicx}	
\usepackage{amsmath}	
\usepackage{amssymb}	
\usepackage{siunitx}    
\usepackage{bbm} 		
\usepackage{xcolor}		
\usepackage{soul}		
\usepackage{booktabs}	
\usepackage{enumitem}	

\title[]{Temporal clustering of rotational glitches in the Crab pulsar}

\author[Carlin et al.]{
J. B. Carlin,$^{1}$
A. Melatos,$^{1,\,2}$
and D. Vukcevic$^{3,\,4}$
\\
$^{1}$School of Physics, University of Melbourne, Parkville, VIC 3010, Australia\\
$^{2}$Australian Research Council Centre of Excellence for Gravitational Wave Discovery (OzGrav)\\
$^{3}$School of Mathematics and Statistics, University of Melbourne, Parkville, VIC 3010, Australia\\
$^{4}$Melbourne Integrative Genomics, University of Melbourne, Parkville, VIC 3010, Australia
}

\date{Accepted XXX. Received YYY; in original form ZZZ}

\pubyear{2018}

\begin{document}
\label{firstpage}
\pagerange{\pageref{firstpage}--\pageref{lastpage}}
\maketitle

\begin{abstract}
It is an open question whether glitch activity in individual pulsars varies on decadal time-scales. The Crab pulsar has experienced 23 spin-up glitches in the last 36 years, interrupting an otherwise monotonic deceleration. A homogeneous Poisson process, i.e. a process with constant rate, is not sufficient to describe the time-ordered distribution of glitch epochs in the Crab pulsar. There are signs of clustering at the $2\sigma$ level when testing with Ripley's $K$ function. Two alternative, inhomogeneous models with one and two step-wise rate changes are found to have higher relative evidence (Bayes factors of 1.74 and 2.86 respectively) than the homogeneous Poisson process. The distinction between clustering, where events are correlated, and rate variation is discussed. The implications for glitch microphysics, in particular trigger mechanisms based on avalanche processes, are briefly discussed.
\end{abstract}




\section{Introduction}
\label{sec:intro}

Rotation-powered pulsars spin down monotonically as they age due to
electromagnetic braking \citep{Taylor1993}. The secular spin down is perturbed
by continuous, stochastic ``timing noise'' \citep{Cordes1980, Arzoumanian1994,
Hobbs2010} as well as impulsive spin-up events called ``glitches''
\citep{Melatos2008, Espinoza2011, Fuentes2017}. Glitch rates and sizes appear to be
uncorrelated in most pulsars \citep{Melatos2018}, with PSR J0537$-$6910 being a
notable exception \citep{Middleditch2006, Ferdman2017}. However the datasets
are small: 531 (386) glitches have been observed in total from 187 (132)
pulsars\footnote{Catalogues of pulsar glitches are maintained independently by
the Jodrell Bank Centre for Astrophysics at
\url{http://www.jb.man.ac.uk/pulsar/glitches.html} and the Australia Telescope
National Facility (ATNF) at
\url{http://www.atnf.csiro.au/research/pulsar/psrcat/glitchTbl.html}. Numbers
in the text without (with) parentheses refer to Jodrell Bank (ATNF) data and
are current as of 2018 September 12.} \citep{Espinoza2011, Yu2013}. Several glitch
trigger mechanisms have been proposed including starquakes \citep{Larson2002,
Negi2007}, superfluid vortex avalanches \citep{Anderson1975, Cheng1988,
Warszawski2011, Warszawski2012}, and magnetospheric phase transitions
\citep{Keith2013} among others; see \citet{Haskell2015} for a modern review.

A systematic study of the clustering of glitch epochs has not been undertaken
across the known population of glitching pulsars, because until recently the
samples per pulsar were too small to reliably infer clustering or rate changes.
Gradually, however, the situation is improving. It is now possible to
disaggregate the data and construct statistically reliable size and waiting
time probability density functions for several objects \citep{Melatos2008,
Howitt2018} and study their size--waiting-time correlations \citep{Fulgenzi2017,
Melatos2018}. Temporal clustering may offer a way to distinguish glitch models based
on vortex avalanches and starquakes. The study of clustering of terrestrial
earthquakes is a large and evolving field. For example, Omori's Law
phenomenologically describes the observed behaviour of aftershocks following a large
earthquake \citep{Omori1894, Utsu1995}. Likewise, with the advent of
Gross-Pitaevskii simulations, it is possible to calculate clustering in superfluid
vortex avalanches theoretically \citep{Warszawski2011, Warszawski2012,
Warszawski2013, Fulgenzi2017}. Neutron star glitches exhibit signs of self-organised
criticality (SOC) \citep{Melatos2008}. Whether all SOC systems show
signs of temporal clustering is not yet clear \citep{Kagan1991, Jensen1998,
Turcotte1999a, Kagan2011, Aschwanden2018}.

An analysis of the Crab pulsar's glitch epochs was carried out by
\citet{Lyne2015}, who claimed that ``the glitches are more clustered than can
be expected from a random occurrence of glitches'' (verbatim quote). In this
paper, we reanalyse the latter data in order to test the hypothesis, that the
Crab's glitches can be modelled as a homogeneous (i.e.\ constant rate) Poisson
process. We review the observations in Section \ref{sec:crab_obs}. In Section
\ref{sec:clustering} we test for clustering in the Crab's glitches using
Ripley's $K$ function \citep{Ripley1977, Ripley1988}, under the null hypothesis
of a homogeneous rate. We emphasize that clustering and rate changes are
different things, as explained in Section \ref{sec:clustering}. We then model
the Crab's glitch activity as a Poisson process that changes rate exactly once
in Section \ref{sec:tworate}. In Section \ref{sec:threerate} we model the
Crab's glitch activity as a Poisson process with two discrete rate changes. We do
a pairwise Bayesian comparison between the one-, two-, and three-rate models in
Section \ref{sec:bayescomp}. 

\section{Observations}
\label{sec:crab_obs}
Daily observations of the Crab pulsar have been carried out by the Jodrell Bank
Observatory since 1982, mainly using the 13m radio telescope at 610 MHz
\citep{Lyne1988, Lyne1993}. These observations are supplemented by earlier,
less regular data from the Arecibo Observatory \citep{Gullahorn1977}, and with
observations in the optical from the Princeton University Observatory
\citep{Groth1975a} and Hamburg Observatory \citep{Lohsen1981}. An overview of
the last 45 years of Crab pulsar observation can be found in \citet{Lyne2015}.

Although 27 glitches have been directly observed, the analysis in \citet{Lyne2015}
is restricted to the 20 glitches that occurred between Modified Julian Date (MJD)
45028 and MJD 55876, as that is when high-cadence monitoring occurred at the Jodrell
Bank Observatory. The cut aims to remove the effect that the 3-yr gap in monitoring
(between 1979 and 1982) may have on the completeness of the dataset. In this paper
we analyse the 23 glitches between MJD 45028 and MJD 58380, as three more glitches
have occurred since 2015, on MJD 57840, MJD 58065, and MJD 58237 \citep{Shaw2018}.
The data from the last 36 years constitutes a statistically complete set, in the
sense specified in section 3.2 of \citet{Espinoza2014}.

\begin{figure}
	\includegraphics[width=\linewidth]{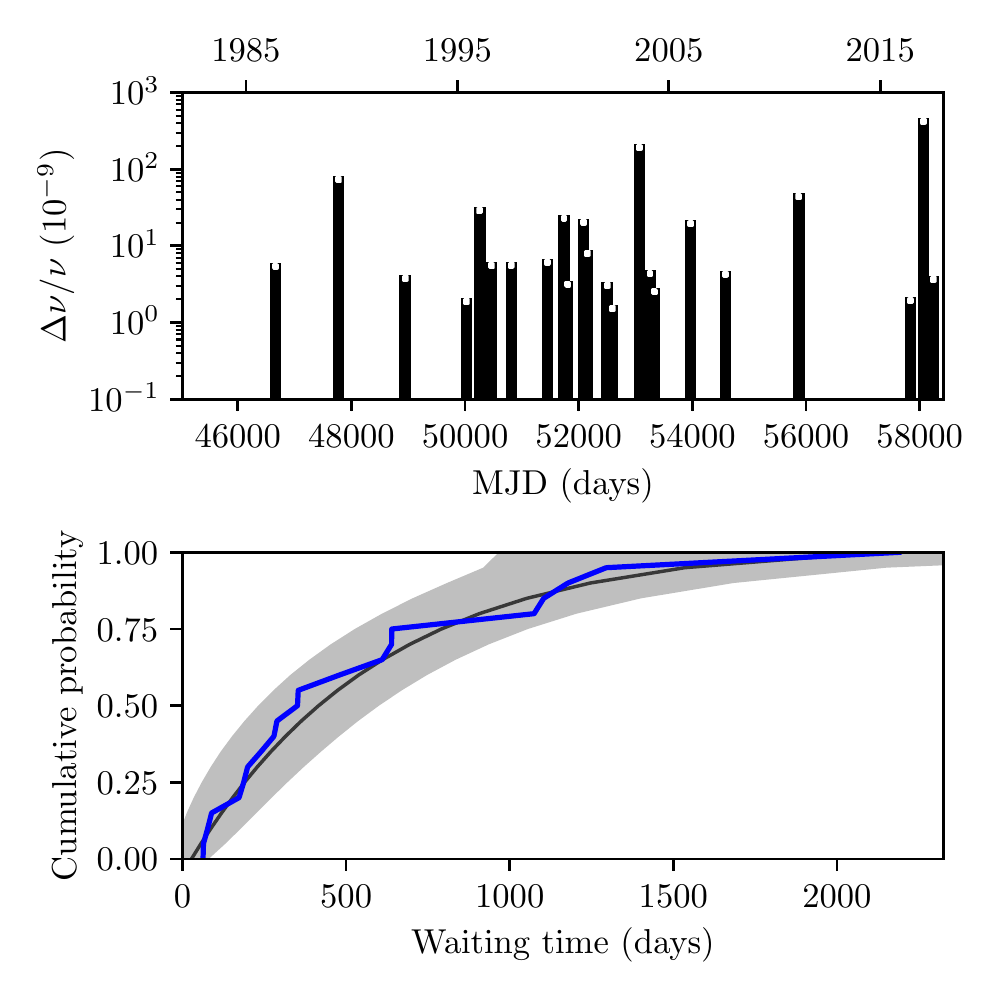}
	\caption{(Top panel.) Epochs and fractional sizes of the Crab's 23
        glitches observed from MJD 45028 (1982) to MJD 58380 (2018). (Bottom
        panel.) Empirical CDF of the 22 observed waiting times between
        consecutive events (blue curve) and for synthetic waiting times
        generated by a homogeneous Poisson process, conditional on 23 events
        occurring in 36 years (black curve, mean; shaded grey region $2\sigma$
        spread; $10^5$ realizations).}
	\label{fig:crab_sizes}
\end{figure}

\subsection{Waiting times}
For the 23 events between MJD 45028 and MJD 58380, the average waiting time between
consecutive glitches is 526.08 days. The epochs, $t$, and fractional sizes ($\Delta
\nu / \nu$, where $\Delta \nu$ is the glitch size, and $\nu$ is the spin frequency),
can be seen in the top panel of Figure \ref{fig:crab_sizes}. Henceforth $t_i$ corresponds to the epoch of the $i$-th glitch. The bottom panel plots
the empirical cumulative distribution function (CDF) for the waiting times between
consecutive glitches (blue curve) as well as a $2\sigma$ envelope of synthetic CDFs
drawn from a Monte Carlo sampling of $10^5$ homogeneous Poisson processes (black
curve, mean; shaded grey region, $2\sigma$ spread), where 23 events occur in 36
years.\footnote{We model a homogeneous Poisson process conditional on observing
exactly 23 events in 36 years by drawing 23 glitch epochs from a uniform
distribution and taking the difference between consecutive epochs to find the
waiting times between glitches. This is not equivalent to drawing 22 numbers from an
exponential distribution with mean equal to the observed average waiting time
\citep{Baddeley2015}.} Waiting times are distributed exponentially in a homogeneous
Poisson process. A Kolmogorov-Smirnov test \citep{Lilliefors1969} does not support
rejecting the null hypothesis, that the 22 inter-glitch waiting times are drawn from
an exponential distribution (p-value of 0.39).

\subsection{Preliminary evidence of clustering}
The Kolmogorov-Smirnov test above involves a summary statistic of the full,
time-ordered dataset of glitch epochs. It only tests one property of the process,
namely that waiting times between consecutive events are distributed exponentially.
It should not be interpreted to imply that a homogeneous Poisson process is the best
or only model for the data.

\begin{figure}
	\centering
	\includegraphics[width=0.9\linewidth]{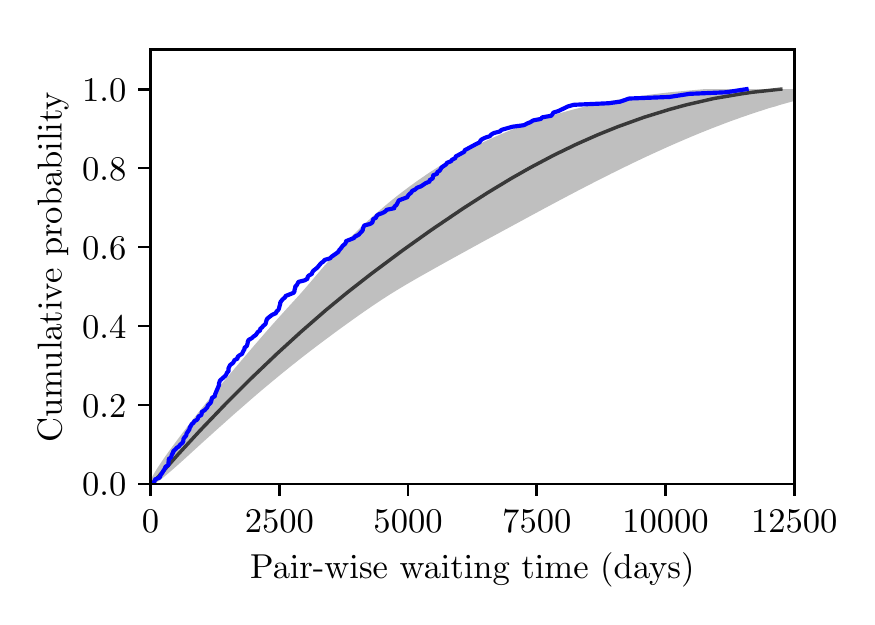}
	\caption{Cumulative distribution function of the 253 waiting times
        between all pairs of glitch epochs for the Crab (blue curve) and for a
        synthetic homogeneous Poisson process, where 23 events occur in 36
        years (black curve, mean; shaded grey region, $2\sigma$ spread; $10^5$
        realizations). [Based on Figure 6 in \citet{Lyne2015}.]}
	\label{fig:cdf_lyne}	
\end{figure}

\citet{Lyne2015} extended this type of test by calculating the empirical CDF of the
waiting times between all pairs of glitch epochs (see Section 3 of the latter
reference). They concluded that a homogeneous Poisson process does not adequately
model the Crab's glitch activity. Figure 6 in \citet{Lyne2015} is reproduced in
Figure \ref{fig:cdf_lyne} of this manuscript. It shows the CDF of 253 all-pair
waiting times, $\Delta t'$. Clustering is indicated if the CDF is steeper at small
$\Delta t'$ than the null model---that the glitch epochs are distributed as a
homogeneous Poisson process---as this shows that there are more events grouped
together than what is expected, if the glitch epochs obey the null model. Figure
\ref{fig:cdf_lyne} shows that the Crab's glitches appear to be clustered, as the
Crab's CDF is 2$\sigma$ above the mean CDF constructed synthetically
from a homogeneous Poisson process at the point of greatest difference. This test
resembles Ripley's $K$ statistic \citep{Ripley1977, Ripley1988} but does not include
an edge correction term, as discussed in the following section.

\section{Clustering}
\label{sec:clustering}
Upon visual inspection, the 14 glitches between MJD 50000 and 53800, in the top
panel in Figure \ref{fig:crab_sizes}, appear to be clustered more closely than the
three glitches between MJD 45028 and MJD 50000, and the six glitches between MJD
53800 and MJD 58380. We now test rigorously whether there is
statistically significant evidence of clustering.

\subsection{Ripley's $K$ function}
An extension of the test used by \citet{Lyne2015} is to use Ripley's \textit{K}
function \citep{Ripley1977, Ripley1988} to test for clustering at a level higher
than what is expected for a homogeneous Poisson process. Ripley's \textit{K}
function is used widely for this purpose in ecology \citep{Kenkel1988,
Peterson1995}, biology \citep{Jafari-Mamaghani2010, Lagache2013}, geoscience
\citep{VegaOrozco2012}, and physics \citep{Loh2008, White2012}. Ripley's \textit{K}
function was applied to one-dimensional spatial data by \citet{Yunta2014} but it
also works for one-dimensional temporal data, such as glitch epochs.

Before we continue we emphasize that a process with clustering is not the same
as a process with a variable rate. For example, in an inhomogeneous Poisson
process, each event is independent, but the underlying rate varies. Often the
underlying cause for the rate change is external and can be measured or
estimated in another way. In contrast, in clustered processes (e.g.\ Cox or
Gibbs), the occurrence of events is correlated, and the underlying rate need
not vary. Often the cause of the clustering is endogenous \citep{Baddeley2015}.
Distinguishing between clustering and an inhomogeneous Poisson process is not
possible from looking at a sequence of events \citep{Bartlett1964,
Baddeley2010}; it is always possible to find a time-varying rate that produces
the same distribution of epochs as a cluster model.

Ripley's \textit{K} function calculates the over- or under-occurrence of
clusters of events in an ordered sequence. It compares to a null hypothesis of
what is expected from a homogeneous Poisson process. Precisely, Ripley's
\textit{K} function counts the number of events that occur within a sliding
window of length $w$,
\begin{equation}
\label{eq:k_func}
{K(w, n) = \frac{T}{n(n-1)} \sum_{i \neq j} H \left(w - \Delta t_{ij} \right) e_{ij}\ \ ,}
\end{equation}
with $\Delta t_{ij} = \mid t_i - t_j \mid $. In \eqref{eq:k_func}, $n$ is the total
number of events in $0\leq w \leq T$. $H()$ is the Heaviside function. The sum
counts how many pairs are composed of events that occur within $w$ of each other,
weighted by an edge correction factor $e_{ij}$. The edge correction we use here is a
variant of Ripley's correction \citep{Ripley1988} for one-dimensional data [see
\citet{Yunta2014} for details]:
\begin{equation}
{e_{ij} = \frac{1}{2} \left[ g(t_i, t_j) + g(t_j, t_i)\right]\ \ ,}
\end{equation}
with
\begin{equation}
{g(t_i, t_j) = 1 + H \left[ \Delta t_{ij} - \text{min}(t_i,\,T - t_i) \right] \ \ .}
\end{equation}
That is, events that occur closer to each other in time than to either edge of
the observation window are weighted more heavily. This accounts for the fact
that edge events may have more neighbours outside the observation window, which
are not observed.

The theoretical $K$ function in one dimension, with the edge correction
specified above, has the following mean and variance \citep{Yunta2014}:
\begin{equation}
\label{eq:k_e}
{\mathbb{E} [K(w, n)] = 2w \ \ ,}
\end{equation}
and 
\begin{equation}
\label{eq:k_var}	
{\text{var}[K(w, n)] = \frac{4 T}{n(n-1)} \left[ w - \frac{5w^2}{4T} + \frac{(n-2)w^3}{6T^2} \right] \ \ .}
\end{equation}
Given \eqref{eq:k_e} and \eqref{eq:k_var} we can introduce a normalised
statistic $\tilde{K}(w,n)$ with zero mean and unit variance,
\begin{equation}
{\tilde{K}(w, n) = \frac{K(w, n) - \mathbb{E} [K(w, n)]}{\sqrt{\text{var}[K(w, n)]}} \ \ .}
\end{equation}

\begin{figure}
	\centering
	\includegraphics[width=0.9\linewidth]{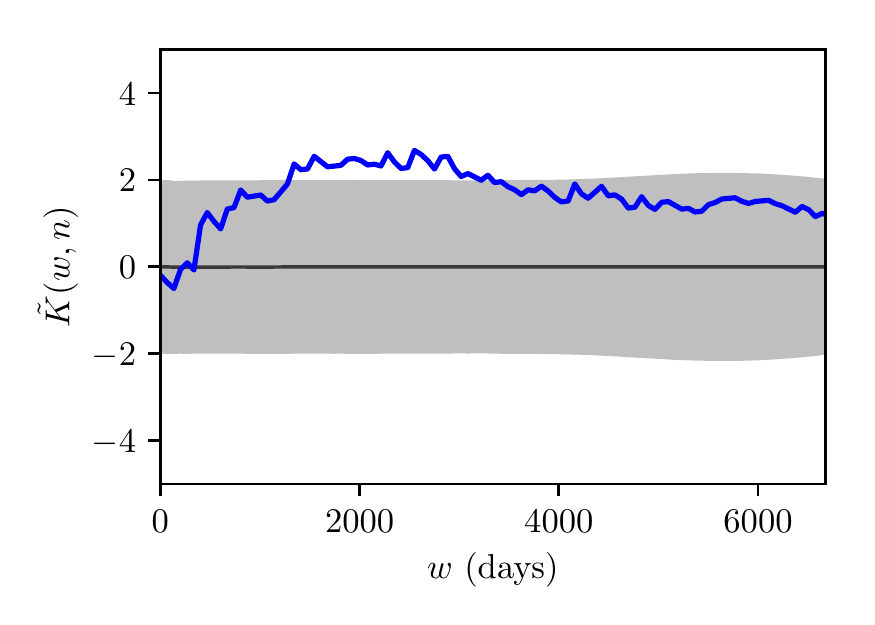}
	\caption{Empirical $\tilde{K}$ function for the Crab (blue curve),
        compared to a synthetic, homogeneous Poisson process, where 23 events
        occur in 36 years (black curve, mean; shaded grey region, $2\sigma$
        spread; $10^5$ realizations).}
	\label{fig:k_stat}
\end{figure}

\subsection{Application to the Crab pulsar's glitches}
The probability density function (PDF) of $\tilde{K}$ is a normal distribution in the
limit $n\gg1$ \citep{Lang2010, Yunta2014}. For intermediate and small values of
$n$, the probability density function is not known analytically. However
quantiles can be calculated from Monte Carlo simulations. Hypothesis testing is
performed by comparing the empirical $\tilde{K}$ statistic for the Crab to the
$\tilde{K}$ statistic for the null hypothesis of a homogeneous Poisson process.
The homogeneous Poisson process is the natural null for this statistic.
Explicitly, we use a Monte Carlo method to generate 23 glitch epochs from the
null homogeneous Poisson process model $10^5$ times, find $\tilde{K}$ for each
simulation, and compare the spread of $\tilde{K}$ with the empirical
$\tilde{K}$ statistic for the Crab's 23 glitch epochs. 

The results of the above test are shown in Figure \ref{fig:k_stat}. The empirical
$\tilde{K}$ statistic for the Crab excedes the $2\sigma$ envelope of the
simulations for 1400 days $\lesssim w \lesssim$ 3000 days. Hence we can say that, under the assumption of a homogeneous Poisson process,
the Crab's glitches exhibit clustering above $2\sigma$ significance for time-scales 1400 days $\lesssim w \lesssim$ 3000 days. This is qualitatively concordant with
the results of the test done by \citet{Lyne2015}, where the empirical statistic for
the Crab (in their case the CDF of waiting times between all pairs of glitches) sits
outside the $2\sigma$ simulation envelope of the null homogeneous Poisson process,
for all-pair waiting times between $\num{1.4e3}$ and $\num{8.0e3}$ days; see Figure
6 in \citet{Lyne2015} and Figure \ref{fig:cdf_lyne} in this manuscript.

\section{Poisson process with one discrete rate change}
\label{sec:tworate}

\begin{figure}
	\includegraphics[width=\linewidth]{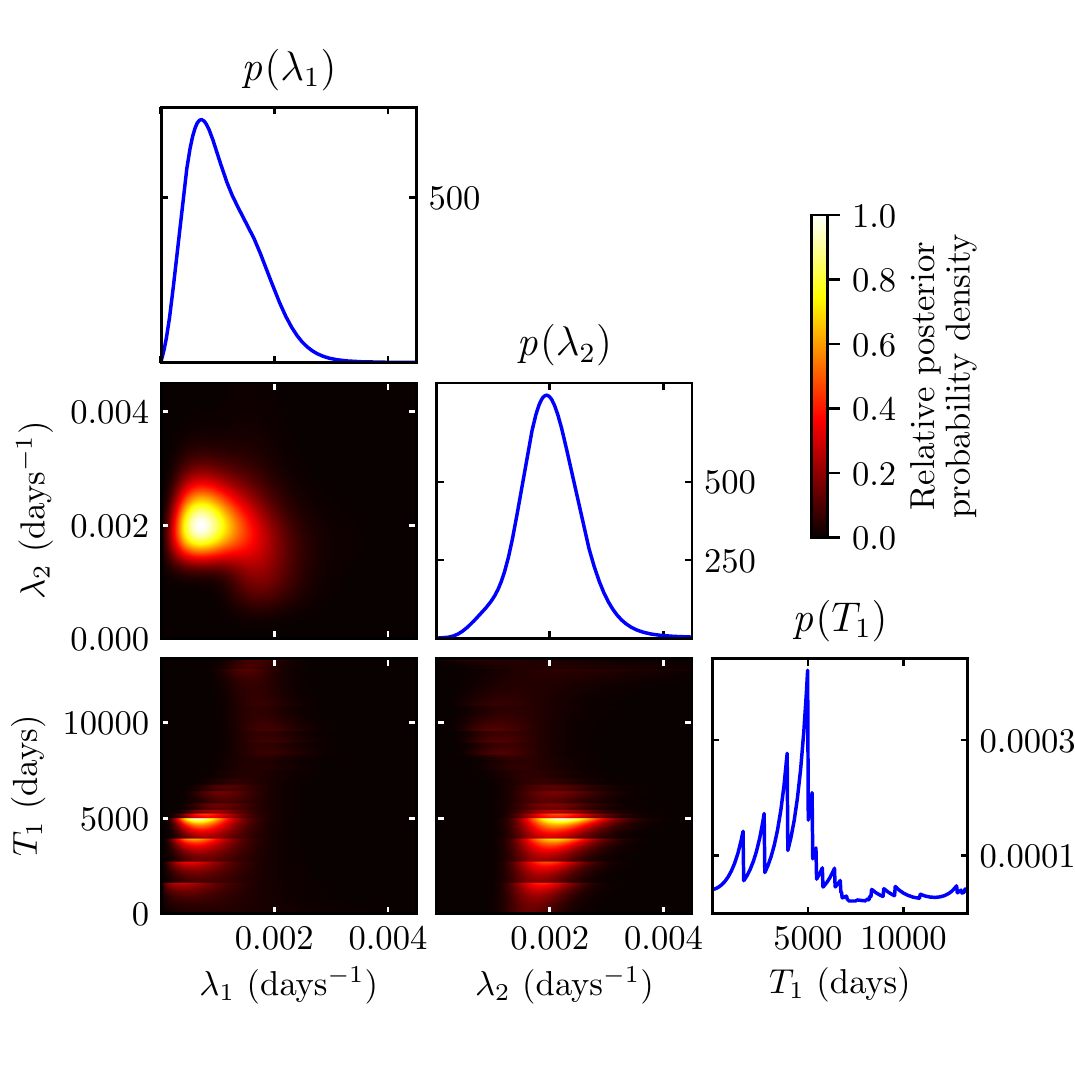}
	\caption{Posterior PDF, $\Pr(\lambda_1, \lambda_2, T_1  \mid  D)$, for the piecewise homogeneous Poisson process with a single rate change at the epoch $T_1$, marginalized over all combinations of parameters. The bottom-left panel is the posterior marginalized over $\lambda_2$, the bottom-middle panel is the posterior marginalized over $\lambda_1$, and the middle-left panel is the posterior marginalized over $T_1$. Single-variable marginal posteriors are also plotted for $\lambda_1$ (top-left), $\lambda_2$ (middle), and $T_1$ (bottom-right). In the two-variable panels the maximum probability point is scaled arbitrarily to unity.}
	\label{fig:newlikely_tworate}
\end{figure}

An exponential waiting-time PDF, which is characteristic of a homogeneous (i.e.\
constant rate) Poisson process, is an adequate fit for the 22 waiting times in
Figure \ref{fig:crab_sizes}. However there are many other ways to get an
approximately exponential waiting-time PDF, for example a state-dependent Poisson
process \citep{Fulgenzi2017}, which is inhomogeneous. The evidence of clustering
found in Section \ref{sec:clustering} indicates that a homogeneous Poisson process
is not sufficient to model the specific, time-ordered sequence of glitch epochs. It
is therefore interesting to ask what simple inhomogeneous models could reasonably
explain the data. We first explore a piecewise homogeneous Poisson process with a
single, instantaneous rate change at the epoch $T_1$. We propose this model because
it is simple, inhomogeneous, and can be extended easily by adding additional rate
jumps.

\subsection{Joint probability density}
To test for a single rate change, we note that the waiting times, $\Delta t_i$,
between consecutive glitches are distributed exponentially for a Poisson
process, i.e.\ the probability density function for the waiting time $\Delta
t_i$ is
\begin{equation}
\label{eq:f_exp}
{	f(\Delta t_i) = \lambda_i e^{-\lambda_i \Delta t_i}}\ \ .
\end{equation}
where $\lambda_i$ is the rate (assumed constant) in the interval between the two
glitches, at epochs $t_i$ and $t_{i+1}$. The complementary CDF, $F(\Delta t_i)$,
which gives the probability of not seeing an event within an interval of length
$\Delta t_i$, takes the form
\begin{equation}
	{F(\Delta t_i) = 1 - e^{-\lambda_i \Delta t_i}}\ \ .
\end{equation}

Whenever the rate changes, there is a time period between two glitches
that brackets the rate change, in which \eqref{eq:f_exp} does not apply. As the
Poisson process is memoryless we can split up each interval between
consecutive glitches into two sections, one of length $\tau_1$ with rate
$\lambda_1$, the other of length $\tau_2$ with rate $\lambda_2$. The probability
density of observing a waiting time of length $\Delta t_i$ is the product,
\begin{align}
	{f(\Delta t_i) =} &\ {\left[1 - F(\tau_1)\right]\ f(\tau_2)} \\
	= &\ {\lambda_2 e^{-\lambda_1 \tau_1 - \lambda_2 \tau_2}}\ \ ,
\end{align}	
with $\Delta t_i = \tau_1 + \tau_2$. If the rate does not change within $\Delta
t_i$ (i.e.\ $\lambda_1 = \lambda_2$), we recover $f(\Delta t_i)$ in
\eqref{eq:f_exp}. 

If we allow the rate to change once only, at the epoch $T_1$, during the total
observation of duration $T$, we can construct the joint probability of seeing the
observed sequence of glitch epochs, $D=\{t_1,\ \ldots\,, t_n\}$, where there are
$N_1$ glitches before $T_1$, and $N_2$ glitches after $T_1$, by multiplying together
the probability density for each consecutive waiting time $\Delta t_i$:

\begin{align}
	{\Pr(D  \mid  \lambda_1, \lambda_2, T_1) =} &\ { \prod_{i=1}^n f(\Delta t_i)} \\ 
	= &\ {\lambda_1^{N_1} e^{-\lambda_1 T_1} \lambda_2^{N_2} e^{-\lambda_2 (T - T_1)}}\ \ .
	\label{eq:exp_prob}
\end{align}

A discussion of an alternative, seemingly natural, yet inappropriate likelihood
function can be found in Appendix \ref{sec:wrong_p}.

\subsection{Marginalized posteriors}
The joint probability distribution \eqref{eq:exp_prob} is a likelihood function,
$L(\lambda_1, \lambda_2, T_1 \mid D)$, i.e.\ the probability of the data, $D$, given
the parameter vector $(\lambda_1,\,\lambda_2,\,T_1)$. We assume a bounded uniform
prior on $T_1$, and gamma distribution priors for $\lambda_1$ and $\lambda_2$ to
calculate the posterior probability density function, $\Pr(\lambda_1,\,\lambda_2,\,
T_1 \mid D)$, for these parameters. The range allowed for $T_1$ covers the entire
observation, i.e.\ $0 \leq T_1/(1\,\text{day}) \leq \num{1.34e4}$. The gamma
distribution PDF
\begin{equation}
\label{eq:gamma}
\textit{p}(\lambda;\, k, \theta) = \frac{1}{\Gamma(k)\theta^k} \lambda^{k -1}
e^{-\lambda/\theta}\ \ ,
\end{equation}
where $\Gamma(k)$ is the gamma function, has shape parameter $k=2$ and scale
parameter $\theta = (N_1 + N_2)/(kT) \approx \num{8.61e-4}\,\text{days}^{-1}$, such
that the mean equals the average rate observed over $0 \leq t \leq T$. The impact of
the prior on the posterior and the Bayesian evidence is explored further in Section
\ref{sec:bayescomp} and Appendix \ref{sec:sens}. There are many valid ways to choose
the prior, of course. Here we adopt the gamma distribution defined in
\eqref{eq:gamma}, which is conjugate to the Poisson distribution in the sense
described in Appendix \ref{sec:sens}. Conjugate priors are commonly adopted in
statistics in the absence of more specific prior knowledge. A bounded uniform prior
for the rates is also defensible but it places a high probability for certain
outcomes that are known to be a priori implausible by eyeballing the data, e.g.\
two simultaneously high rates. In contrast, the gamma distribution defined in
\eqref{eq:gamma} decreases to zero as $\lambda$ increases, which is more realistic.
Hence we prefer the gamma prior for the rates in this paper.

We can marginalise the full posterior probability, $\Pr(\lambda_1,\,\lambda_2,\, T_1
\mid D)$, over any one parameter to find the marginalized posterior probability of
the other two. The results of marginalising can be seen in Figure
\ref{fig:newlikely_tworate}. The two-dimensional heat maps (middle-left,
bottom-left, and bottom-middle panels) show the joint likelihood of different pairs
of parameters. The bottom-right panel indicates that a model with a rate change just
before $T_1 \approx \num{5e3}$ days has the best support. This value of
$T_1$ corresponds roughly to MJD 50000. In the middle-left panel, much of the
probability mass lies off the diagonal, suggesting that there is little support for
a homogeneous (constant rate) Poisson process. The marginalized posterior for $T_1$
is spiky, because \eqref{eq:exp_prob} changes discontinuously when $T_1$ rolls over
a glitch epoch, which increments $N_1$ and decrements $N_2$. In contrast, the
marginalized posteriors for $\lambda_1$ and $\lambda_2$ are smooth, because $N_1$
and $N_2$ are not functions of $\lambda_1$ and $\lambda_2$.

\begin{figure*}
	\includegraphics[width=0.75\textwidth]{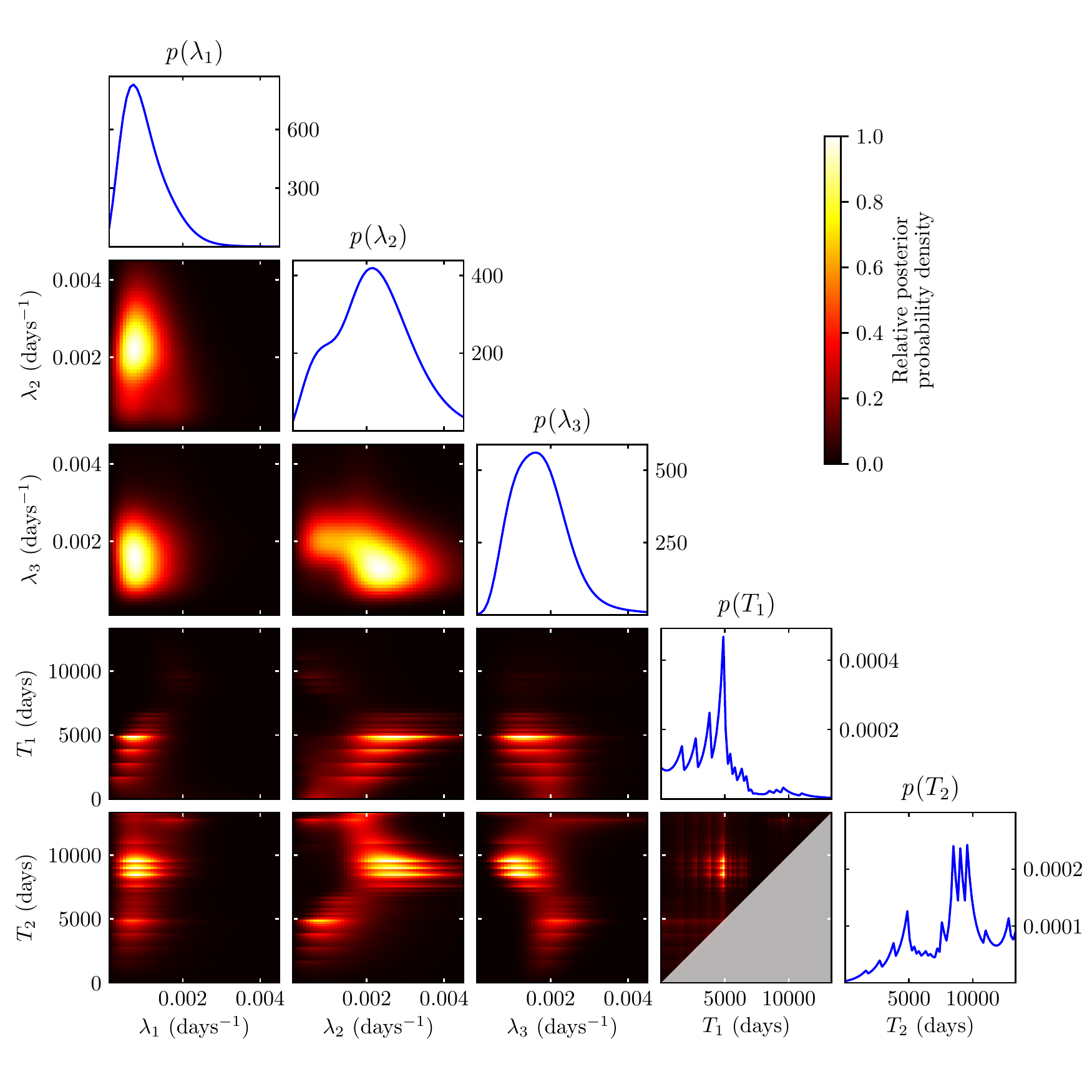}
	\caption{Posterior PDF, $\Pr(\lambda_1, \lambda_2, \lambda_3, T_1, T_2  \mid  D)$, for the piecewise homogeneous Poisson process with two rate changes, one at the epoch $T_1$, the other at the epoch $T_2$, marginalized over all combinations of parameters. In the two-variable panels the maximum probability point is scaled arbitrarily to unity.}
	\label{fig:marg_threerate}
\end{figure*}

\section{Poisson process with two discrete rate changes}
\label{sec:threerate}
We can also test whether the Crab's glitches can be modelled as a piecewise
homogeneous Poisson process with two instantaneous rate changes, one at the
epoch $T_1$, the second at the epoch $T_2 > T_1$. This more elaborate model is
inspired by the top panel of Figure \ref{fig:crab_sizes}, where there appear
(to the eye) to be three distinct regimes of glitch activity: before the glitch
at MJD 50000, between MJD 50000 and MJD 53800, and after MJD 53800. We test for
the two-rate-change model by extending \eqref{eq:exp_prob} to the joint
probability of observing the sequence of glitch epochs, $D$, where there are
$N_1$ glitches in $0 \leq t < T_1$, $N_2$ glitches in $T_1 \leq t < T_2$, and
$N_3$ glitches in $T_2 \leq t \leq T$, with constant rates $\lambda_1$,
$\lambda_2$, and $\lambda_3$ respectively. The result is

\begin{align}
\label{eq:prob_3rate}
{\Pr(D \mid  \lambda_1, \lambda_2, \lambda_3, T_1, T_2)} =&\ \lambda_1^{N_1} \textit{e}^{-\lambda_1 T_1} \lambda_2^{N_2} \textit{e}^{-\lambda_2 (T_2 - T_1)} \nonumber \\
&\times \lambda_3^{N_3} \textit{e}^{-\lambda_3 (T - T_2)} \ \ .
\end{align}

The joint probability defined by \eqref{eq:prob_3rate} is a likelihood
function, as in Section \ref{sec:tworate}, i.e.\ the probability of the
data, $D$, given the parameter vector $(\lambda_1,\ \lambda_2,\ \lambda_3,\
T_1,\ T_2)$. We assume the same gamma distribution prior defined in Section
\ref{sec:tworate} on all three rate parameters $\lambda_1$, $\lambda_2$, and
$\lambda_3$. For $T_1$ and $T_2$ we assume a uniform prior on the
space of all possible values, which is the triangle (two-dimensional simplex)
defined by the natural bounds on $T_1$ and $T_2$, including $T_1 < T_2$.  This
triangle is depicted as the non-grey portion of the $T_1$-$T_2$ panel in Figure
\ref{fig:marg_threerate}. 

The marginalised posterior PDF is shown in Figure
\ref{fig:marg_threerate}. The panels along the diagonal from the top-left to
the bottom-right show the one-dimensional marginalised posteriors for
$\lambda_1$, $\lambda_2$, $\lambda_3$, $T_1$, and $T_2$ respectively. The sharp
peak in $\textit{p}(T_1)$ at $T_1 \approx \num{5e3}$ days corresponds to roughly MJD
50000. The three sharp peaks in $\textit{p}(T_2)$ between $\num{8e3}$
days and $10^4$ days correspond to the three glitches following MJD
53800. In other words, when using two rate-change model, the data most
supports one rate change occurring at around MJD 50000 and a second rate change
occurring somewhere between MJD 53800 and MJD 54000. Broadly speaking, as the
marginalised posteriors for $T_1$ and $T_2$ are approximately unimodal there is
qualitative support for two rate changes fitting the Crab's glitches. However, this
statement is purely qualitative. We assess its worth quantitatively in Section
\ref{sec:bayescomp}, through a formal model comparison using Bayes factors.

\begin{figure}
	\centering
	\includegraphics[width=0.85\linewidth]{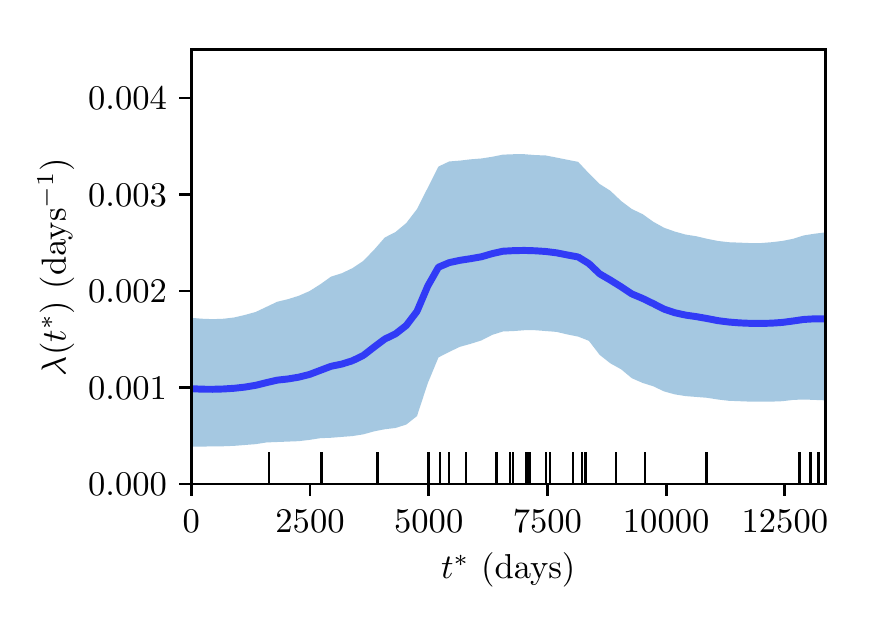}
	\caption{The solid blue curve shows the mean of the PDF $p[\lambda(t^*)]$ of $\lambda(t^*)$. The light blue band indicates the 10th and 90th percentile of each PDF. The small black ticks indicate the epochs of the 23 observed glitches.}
	\label{fig:lam_tstar}
\end{figure}

Another way to represent the posterior distribution is to pick an observation time,
$t^*$, and construct a PDF for the value of the rate at that time, $\lambda(t^*)$.
As this rate function is a piecewise constant function of the parameters in the
model, the posterior distribution of $\lambda(t^*)$ is an integral of the full
posterior distribution on the five parameters, $(\lambda_1, \lambda_2, \lambda_3,
T_1, T_2)$. $\lambda(t^*)$ is a function of the posterior PDF, and as such uses all
of the data, before and after $t^*$. We use a grid-based numerical integration
method. The result is shown in Figure \ref{fig:lam_tstar}. This figure divides into
roughly three regimes. $t^*\leq \num{5e3}$ days, where $\lambda(t^*)
\approx 0.001$ days$^{-1}$; $5 < t^*/(10^3\,\text{days}) \leq 8$, where
$\lambda(t^*) \approx 0.0025$ days$^{-1}$; and $t^* >
\num{8e3}$ days, where $\lambda(t^*) \approx 0.0015$
days$^{-1}$. We emphasise that the presence of three regimes is not surprising,
because we are fitting a two-rate-change model. Nevertheless, this summary statistic
derived from full posterior offers a way to visualize the
uncertainty in the estimate of the rate function, given the choice of a
two-rate-change model. The relative Bayesian evidence in the data for this model is
calculated in Section \ref{sec:bayescomp}.

\section{Bayesian Model Selection}
\label{sec:bayescomp}
To quantitatively compare the relative goodness-of-fit for the homogeneous
Poisson process (model $M_0$), one rate change Poisson process ($M_1$), and two
rate change Poisson process ($M_2$), we compute the Bayesian evidence for
$M_0$, $M_1$ and $M_2$ by integrating the likelihood function over all
parameters, viz. 
\begin{equation}
	\label{eq:ev}
	\text{Evidence}(M_i  \mid  D) = {\int d\vec{\theta}_i\ L(M_i, \vec{\theta}_i \mid D)\ \Pr(\vec{\theta}_i)}\ \ ,
\end{equation}
where $\Pr(\vec{\theta}_i)$ are the priors for the parameter vectors $\vec{\theta}_i$.
Explicitly, the priors chosen for all models $M_i$ are
\begin{align}
	\Pr(\lambda_1) =&\ \text{Gamma}\,(\lambda_1;\,k=2,\,\theta=\num{8.61e-4}\,\text{days}^{-1})\ \ ,
\end{align}
and $\Pr(\lambda_1) = \Pr(\lambda_2) = \Pr(\lambda_3)$. For $M_1$, we also have
\begin{align}
	\Pr(T_1) = &\ U_1 \, (0\text{ days} \leq T_1 \leq \num{1.34e4}\text{ days})\ \ .
\end{align}
For $M_2$, we also have
\begin{align}
    \Pr(T_1, T_2) = &\ U_2 \, (0\text{ days} \leq T_1 < T_2 \leq \num{1.34e4}\text{ days})\ \ ,
\end{align}
where $U_1$ and $U_2$ are normalized to unity. $\text{Gamma}\,(\lambda;\,k,\,\theta)$ refers to the gamma distribution defined in \eqref{eq:gamma}.

\begin{table}
	\centering
	\begin{tabular}{cc}
		\toprule
		Model & Evidence ($\times10^{-75}$) \\
		\midrule
		$M_0$ & \phantom{1}{7.56} \\
		$M_1$ & {13.14} \\
		$M_2$ & {21.62} \\
		\bottomrule
	\end{tabular}
	\caption{Bayesian evidence for the homogeneous Poisson process (model $M_0$),
one rate change Poisson process ($M_1$), and two rate change Poisson process
($M_2$).}
	\label{tab:bf_unif}
\end{table}

The integrals in \eqref{eq:ev} are performed using adaptive quadrature
\citep{Piessens1983}. The results can be seen in Table \ref{tab:bf_unif}. The
absolute scale is not important for model selection. To compute the Bayes factor we
take the ratio of the evidence for any two models. The ratio provides a scale on
which to judge the relative evidence in the data for two competing models. The
largest Bayes factor is for $M_2$ over $M_0$ (Bayes factor 2.86). This is
interpreted as ``positive'' or ``substantial'' evidence, depending on which standard
scale one prefers \citep{Kass1995, Jeffreys1998}. Bayesian evidence and
Bayes factors automatically control for over-fitting (i.e. too many model
parameters) by explicitly recognising and integrating over the prior distributions.
If the likelihood of the data given two different models is the same, but one model
has a larger prior parameter space, then it will have lower relative evidence
compared to the other model \citep{Mcelreath2016}. Broadly speaking, both
inhomogeneous rate models $M_1$ and $M_2$ are favored over the null hypothesis of a
homogeneous Poisson process ($M_0$), given the priors specified above. The
sensitivity of the evidence to the parameters in the gamma distribution prior is
explored in Appendix \ref{sec:sens}.

\section{Conclusions}
\label{sec:concl}
Visual inspection of the epochs for the 23 Crab pulsar glitches between 1982
and 2018 suggests that they may be clustered and/or occur at a variable rate,
with the 14 glitches between MJD 50000 and MJD 53800 seeming to occur more
frequently than the rest. This visual observation was tested quantitatively by
\citet{Lyne2015}, who showed that the CDF for the
waiting times between all glitch pairs cannot be modelled adequately as a
homogeneous Poisson process. In this paper we look at the question from two
perspectives, which complement the approach in \citet{Lyne2015}. Specifically,
(i) we test for clustering using Ripley's $K$ function, and (ii) we test the
support for two alternative, inhomogeneous Poisson processes. 

If we assume a homogeneous Poisson process, there are signs of clustering at
$2\sigma$ significance for time-scales 1400 days $\lesssim w \lesssim$ 3000 days, as indicated by Ripley's $K$ function (Section
\ref{sec:clustering}). Assuming gamma distribution priors for the
rates, inhomogeneous Poisson processes tested with one and two
discrete rate changes are both favoured over a homogeneous Poisson processes, with
Bayes factors of 1.73 and 2.86 respectively (Sections
\ref{sec:tworate}--\ref{sec:bayescomp}). 

Although the glitch trigger mechanism is unknown, many discussions of the issue
focus on avalanche processes, involving superfluid vortex motion or starquakes for
example. In an avalanche process one must distinguish between two rates: (i) the
rate at which the system is driven, which is set externally and is usually slow and
constant; and (ii) the event rate, which is determined internally by the avalanche
dynamics and fluctuates around a mean value. Therefore sequences of events observed
during a short block of time (short compared to the time-scale of the external
driver) can be interpreted in terms of rate changes or clustering; see Figure 24 in
\citet{Fulgenzi2017}. This result equally applies to starquakes and superfluid
vortex avalanches if the underlying processes involve a large number of interacting
elements (e.g. tectonic plates, vortices) engaged in correlated, ``knock-on'' motion
\citep{Haskell2015}, e.g. as in SOC systems \citep{Jensen1998}. Processes that do
not involve avalanches can also exhibit rate changes, e.g. temperature changes or a
build up of differential rotation, but it is unclear why these would change markedly
on decadal time-scales, and do so non-monotonically; see Figure \ref{fig:lam_tstar}.

Future work along these lines includes: (i) testing whether the data are
Markovian, i.e.\ that a glitch waiting time depends solely on the previous waiting
time \citep{Cox1966, Cinlar1975}; (ii) treating the data as a marked Poisson
process by introducing the additional information of glitch sizes
\citep{Kingman1993}; or (iii) applying the physically motivated state-dependent
Poisson process proposed by \citet{Fulgenzi2017}. Model fitting of any kind is
greatly assisted by using all of the information available in the time-ordered
sequence of events observed.

\section{Acknowledgements}
\label{sec:ackn}
Parts of this research were conducted by the Australian Research Council Centre
of Excellence for Gravitational Wave Discovery (OzGrav) through project number
CE170100004. The authors would like to thank John Bowman at Walmart Labs (user
``jbowman'' on \url{stats.stackexchange.com}) for suggesting useful
modifications to the likelihood function.


\appendix
\section{Alternative joint probability distribution}
\label{sec:wrong_p}
Instead of equation \eqref{eq:exp_prob} one may be tempted to construct a different
likelihood function of the form
\begin{align}
{\Pr(N_1, N_2  \mid \lambda_1, \lambda_2, T_1)}\ =&\ {\frac{\left(\lambda_1 T_1 \right)^{N_1} \textit{e}^{-\lambda_1 T_1}}{N_1 !}} \nonumber\\
		& \times\ \frac{\left(\lambda_2 (T - T_1) \right)^{N_2} \textit{e}^{-\lambda_2 (T - T_1)}}{N_2 !}\ \ ,
\label{eq:wrong_p}
\end{align}
because this is the joint probability of one Poisson distribution with $N_1$ events
in a region of length $T_1$, at a constant rate $\lambda_1$, multiplied by another
Poisson distribution with $N_2$ events in a region of length $(T - T_1)$, at a
constant rate $\lambda_2$. However, \eqref{eq:wrong_p} throws away most of the
information about the timing of the glitches. It effectively splits the entire
time-ordered dataset into only two samples, the sets of events before and after
$T_1$, and summarises each sample with an unordered count of the events. It
therefore cannot distinguish properly between small $T_1$ with high $\lambda_1$, or
large $T_1$ with low $\lambda_1$; both scenarios have the same probability in
\eqref{eq:wrong_p}. In contrast, equation \eqref{eq:exp_prob} assigns different
probabilities to these scenarios. Hence \eqref{eq:wrong_p} is not the appropriate
likelihood function to use when asking questions about the timing of events, which
we need to do when testing whether the data can be modelled as an inhomogeneous
Poisson process with one or two discrete rate changes.

\section{Sensitivity of Bayesian evidence to the prior}
\label{sec:sens}
The gamma distribution assumed as the prior for the rates has the functional form
given in \eqref{eq:gamma}. The gamma distribution is chosen because it is flexible
and conjugate to a Poisson distribution, i.e.\ the likelihood multiplied by the prior
is of the same functional form as the prior itself \citep{Gelman2013}. This allows
for integrals over the likelihood and prior to be computed efficiently. Prior
observational knowledge about the rate of events for the Crab pulsar, assuming a
homogeneous Poisson process, indicates that the mean of the prior distribution
should be comparable to the mean rate for Crab, i.e.\ $\bar{\lambda} = 23 / (58380 - 45028) \approx \num{1.72e-3}$ days$^{-1}$. The mean of \eqref{eq:gamma} is
$k\theta$, so as our default choice we set $\theta = \bar{\lambda} / k$ in Sections
\ref{sec:tworate}--\ref{sec:bayescomp}. To test the sensitivity of the Bayesian
evidence to this choice we test two alternative values of the mean, namely $k\theta
= 0.5\bar{\lambda}$ and $k\theta = 2\bar{\lambda}$. We also test the sensitivity to
changes in $k$. The Bayes factors for a few representative choices of these
parameters are shown in Table \ref{tab:gamma_sens}.

In all nine parameter sets, the model with two rate changes, $M_2$, has the highest
evidence. The Bayes factors do not vary wildly from set to set. Broadly speaking,
the evidence in favor of the two rate change model is consistent, regardless of the
exact specification of the prior. This gives some confidence that the model truly
captures an essential feature of the data.

\begin{table}
	\centering
	\begin{tabular}{ll  rr}
		\toprule
		$k$ & $k\theta/\bar{\lambda}$ & \multicolumn{1}{c}{$M_1 / M_0$} & \multicolumn{1}{c}{$M_2 / M_0$}\\
		\midrule
		1 & 0.5 & 1.75 & 2.05 \\
		1 & 1.0 & 1.70 & 2.54 \\
		1 & 2.0 & 1.27 & 1.75 \\
		\midrule
		2 & 0.5 & 1.74 & 2.01 \\
		2 & 1.0 & 1.74 & 2.86 \\
		2 & 2.0 & 1.09 & 1.55 \\
		\midrule
		4 & 0.5 & 1.53 & 1.62 \\
		4 & 1.0 & 1.64 & 2.61 \\
		4 & 2.0 & 0.82 & 1.00 \\
		\bottomrule
	\end{tabular}
	\caption{Bayes factors for the one rate change Poisson process over the
homogeneous Poisson process ($M_1/M_0$), and the two rate change Poisson process
over the homogeneous Poisson process ($M_2/M_0$), for nine parameter sets $(k,
k\theta/\bar{\lambda})$ defining the gamma distribution rate prior.}
	\label{tab:gamma_sens}
\end{table}



\bibliographystyle{mnras}
\bibliography{crab_bib} 


\bsp	
\label{lastpage}
\end{document}